\documentclass[final,5p,times,twocolumn]{elsarticle}

\usepackage{lineno}%
\usepackage{bm}
\usepackage{epstopdf}
\usepackage{graphicx}
\usepackage{xcolor}
\usepackage{amssymb}
\usepackage{amsthm}
\usepackage{dcolumn}% Align table columns on decimal point
\usepackage{enumerate}% bold math
\usepackage[colorlinks=true]{hyperref}

\journal{Physica A}

\begin{document}

\begin{frontmatter}

\title{Interparticle Correlations in the Simple Cubic Lattice of Ferroparticles:
Theory and Computer Simulations}% Force line breaks with \\

\author[label1]{Anna Yu. Solovyova}
\author[label2,label3]{Andrey A. Kuznetsov}
\author[label1,label2]{Ekaterina A. Elfimova\corref{2}}
\cortext[2]{Corresponding author.}
\ead{ekaterina.elfimova@urfu.ru}
\address[label1]{Ural Mathematical Center, Ural Federal University, 51 Lenin Avenue, Ekaterinburg 620000, Russia}
\address[label2]{Institute of Natural Sciences and Mathematics, Ural Federal University, 51 Lenin Avenue, Ekaterinburg 620000, Russia}
\address[label3]{Physics of Phase Transitions Department,
	Perm State University, 15 Bukireva St., Perm 614990, Russia
}
\begin{abstract}
Anisotropic interparticle correlations in the simple cubic lattice of single-domain ferroparticles (SCLF) are studied using both theory and computer simulation. The theory is based on the Helmholtz free energy expansion like classical virial series up to the second virial coefficient. The analytical formula for the Helmholtz free energy is incorporated in a logarithmic form to minimize the effects of series truncation. The new theoretical approach, including discrete summation over lattice nodes coordinates, is compared critically against the classical virial expansion of the Helmholtz free energy for the dipolar hard sphere fluid; the main differences between the Helmholtz free energy of SCLF and dipolar hard sphere fluid are discussed.  The theoretical results for the Helmholtz free energy, the magnetization, and the initial magnetic susceptibility of the SCLF are compared against Molecular Dynamic simulation data. In all cases, theoretical predictions using logarithmic form of the Helmholtz free energy are seen to be superior, but they only have an applicability range of the effective dipolar coupling constant $\lambda_e < 1.5$. For highest values of $\lambda_e$, the structural transition of the magnetic dipoles in SCLF is observed in Molecular Dynamic simulation. It has been shown that for $\lambda_e \gtrsim 2$, an antiferromagnetic order appears in the system.
\end{abstract}


\end{frontmatter}

%\linenumbers

\section{Introduction}\label{intro}
More than 50 years ago, Resler and Rosensweig \cite{RESLER1974} reported the first synthesis of a stable suspension of magnetic nanoparticles in a magneto-passive carrier liquid, nowadays known as a ferrofluid or magnetic fluid. This was the first step in designing soft magnetic materials which respond to external magnetic fields. The first attempts to describe ferrofluid properties theoretically \cite{Shliomis1974, Rosensweig1998} were based on the theory of an ideal (noninteracting) paramagnetic gas \cite{Langevin1905, Elmore1938}, according to which the  fractional (scalar) magnetization $M_L$ and initial magnetic susceptibility $\chi_L$ of ferrofluids are determined by a simple and convenient Langevin law:
\begin{equation}\label{Lang}
    M_L = \coth \alpha - \frac{1}{ \alpha}, \ \ \ \ \chi_L=\frac{\rho m^2}{3k_BT} ,
\end{equation}

\noindent where $\rho$ is the magnetic particle number concentration and $\alpha = m H/k_B T$ is the Langevin parameter (the relation of the Zeeman interaction energy of the particle magnetic moment $m$  with the external magnetic field $H$ to the thermal energy $k_BT = \beta^{-1}$).

Since then ferrofluids have been the subject of intense scrutiny, with regard to their structure, phase behavior, and dynamics \cite{Ivanov050602, Holm2005, Odenbach2002, Ivanov2018, Lebedev2019, Szalai2015,Vanhaelen2018}. It became clear that a theory based on a single-particle approximation, like expressions (\ref{Lang}), is only valid  for an infinitely diluted suspension, when the interparticle magnetic interactions can be neglected. It was proved that interparticle magnetic correlations greatly increase the static magnetic susceptibility of ferrofluids \cite{Pshenichnikov2005, Szalai2015, Solovyova2016, Solovyova2017}, change the spectrum of its dynamic susceptibility \cite{Sindt2016, Ivanov2016, Lebedev2019}, and lead to the self-assembly of ferroparticles into chains, rings, branched structures, and three-dimensional percolating networks \cite{Wen1999, Mendelev2005}. A vast range of theoretical models has been developed to link the macroscopic properties of ferrofluids to their internal microscopic structure and the interactions between the magnetic nanoparticles. Examples include Weiss’s mean-field theory \cite{Weiss1907, Tsebers1982}, the mean-spherical approximation closure of the Ornstein-Zernike equation \cite{Wertheim1971, MOROZOV1990}, the high-temperature approximation \cite{BUYEVICH1992}, first-order \cite{PSHENICHNIKOV1996} and second-order modified mean-field theories \cite{Ivanov2001}, and the Born-Mayer cluster-expansion theories \cite{Huke2004, Elfimova2013}.

Today, the design of  soft magnetic materials has steadily progressed: it is now possible to embed magnetic particles into polymer matrices, creating magnetoresponsive elastomers and ferrogels. These materials can be remotely aligned and guided under external fields \cite{Filipcsei2007,Raikher2008, Chertovich2010,Zubarev2018,Snarskii2019}, which has opened up broad opportunities for theirs applications in biomedicine and technology \cite{Zubarev201759, Sanchez2017, Borin2007, Becker2017}. Current experimental techniques offer different strategies for the synthesis of magnetoresponsive elastomers and ferrogels. Usually, in ferrogels the polymer matrix is only  weakly cross-linked; therefore, magnetic particles can diffuse through the network and build some agglomerates. Magnetoresponsive elastomers typically consist of a highly cross-linked matrix which is so compact that translations of particles dispersed in the matrix might be hindered, especially when the particles are large. The spatial distribution of particles in a magnetoresponsive elastomer can be either isotropic or anisotropic  (chain-like, plane-like), that depends on the method of preparation \cite{Filipcsei2007,Zhang2008,Kulichikhin2009,Asadi2019}. Isotropic distribution of magnetic particles inside an elastomer are achieved by cross-linking of a polymer melt with well-dispersed magnetic particles without applying  external magnetic ﬁeld. Using 3D printing technologies allows to embed particles in the given regular order \cite{Elder2020}. The material properties significantly depend on the particles' distribution in the material and the particle-matrix relationship \cite{Elfimova2019_Nanoscale}.

Improving the synthesis technology and developing methods for using magnetoresponsive elastomers and ferrogels requires solving fundamental problems related to predicting the behavior of ensembles of magnetic particles in a polymer matrix. Because of the complex microstructure of these systems, their full explicit atomistic modeling unfeasible. Considerable simplifications at the microscopic level should be chosen very carefully in order to attain a proper representation of the experimental system. Computer simulation models usually represent the embedded magnetic particles as beads with point magnetic dipoles, whereas particle-particle and particle-polymer matrix interactions are modelled with different levels of detail \cite{Wood2011, Ivaneyko2011, Ivaneyko2015, Melenev2019, Ryzhkov2017, Minina2018}. Obviously, the more naturally the microscopic details are presented in a simulation model, the more expensive becomes the cost of computations. Another way to study the behavior of the ensembles of magnetic particles in a polymer matrix is the continuum-mechanics approach, which is based on the numerical solution of balance laws in different formulations and Maxwell’s equation for magnetic field \cite{Brigadnov2003, Zubarev2019}.

Theories resulting in {\it explicit analytical expressions}  characterizing accurately the response of the magnetic particle ensemble in a polymer matrix on the external magnetic field still represent a challenge. The development of these theories is based on two main strategies dealing with the description of the internal structure of these systems. The first takes into account only particle-matrix interactions, using, for example, the effective Jeffreys model \cite{Raikher2013, Raikher2010}, the Kelvin model \cite{Rusakov2017}, or assuming that the particles in the matrix are completely immobilized \cite{RAIKHER2014}. In the latter case, the reaction of the magnetic moment to an external field occurs only according to the Néel mechanism. The second strategy includes accounting for particle-particle  dipolar interactions. These models were  developed only for ensembles of immobilized magnetic particles, that correspond to systems with a sufficiently hard polymer matrix. In particular, the explicit expressions for the static magnetization and initial magnetic susceptibility of ensembles of immobilized superparamagnetic spherical particles were determined in \cite{Ivanov2019, Elfimova2019_Nanoscale}; the dynamic magnetic susceptibility of an ensemble of immobilized superparamagnetic particles in a weak, linearly polarized ac magnetic field was studied in \cite{AMBAROV2020}, where the particles' easy magnetization axes were aligned with some given angle to the ac field. In these studies, it was assumed that particles were randomly distributed and fixed in the matrix; therefore, the discrete material properties were not taken into account. The interparticle dipole-dipole interaction was taken into account based on the first order modified mean-field model, in the framework of which the orientation of the magnetic moment of a randomly chosen particle is influenced by an external magnetic field and by the total dipolar field produced by all other magnetic moments.

In this paper, we study the effect of dipole-dipole interparticle interactions on the static thermodynamic and magnetic properties of an ensemble of magnetic particles embedded in a polymer matrix, taking their microscopic discrete structure explicitly into account. We assume that magnetic particles are embedded on the nodes of the regular cubic lattice and can rotate at lattice nodes under the influence of an external magnetic field and as a result of interparticle dipole-dipole interactions; however, particle translational degrees of freedom are turned off. 
This model will be studied theoretically based on rigorous methods of statistical physics as well as with help of computer simulation. Two types of real magnetic composites correspond to model system considered here. In the first case, the particles have a rigid connection with the carrier matrix and they are characterized by a Neel mechanism of magnetic moment relaxation. However, the internal magnetic anisotropy of particles in this case is low and the magnetic moment can freely rotate inside the particle body. Typically, the size of such particles is small and the dipole-field relationship turns out quite weak. In the second case, the particles can rotate in some "caverns" of the host medium and the orientation of the magnetic moment changes according to the Brownian mechanism. Usually such particles have a large size and react strongly with the field. 

This article is organized as follows. In section \ref{sec:basis}, the model system is described, the analytical results for the Helmholtz free energy expansion are derived, and simulation details are summarized. The results of a comparison of the theoretical predictions with simulation data are presented in section \ref{sec:rslt}. Section \ref{sec:cncld} concludes the article.

\section{Model and methods}\label{sec:basis}
\subsection{Model}

\begin{figure}[t!]
\center
\includegraphics[width=0.9\linewidth]{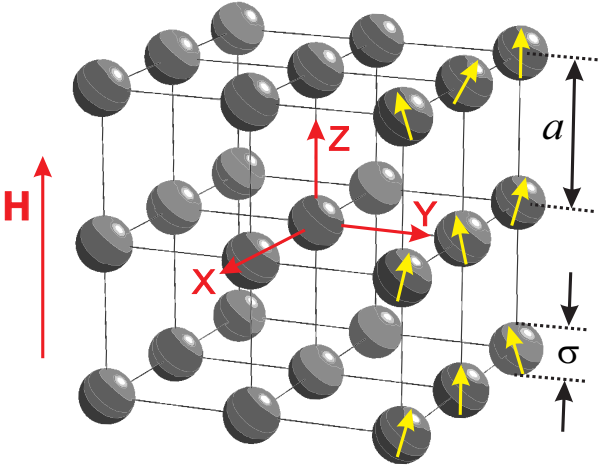}
\caption{Monodisperse system of immobilized single-domain spherical ferroparticles placed at the nodes of a simple cubic lattice.}
\label{fig:model}
\end{figure}

A monodisperse system of $N$ immobilized single-domain spherical ferroparticles placed at the nodes of a simple cubic lattice is considered. Throughout this article, the model system is named as SCLF.
Each ferroparticle has diameter $\sigma$ and magnetic moment $m = \pi \sigma^3 M_s / 6$, where $M_s$ is the bulk saturation magnetization. The simple cubic lattice has the period \emph{a}, which allows us to write the volume \emph{V} of the SCLF as $V= N a^3$. Thus, the number concentration of ferroparticles is defined as $\rho=a^{-3}$. A schematic representation of the SCLF is given in Fig. \ref{fig:model}.

There are no demagnetization fields in both theory and computer simulation. In theory, it is assumed that the system occupies the volume \emph{V}, which has a long cylindrical shape elongated in the direction of external magnetic field \textbf{\textrm{H}}. In the simulation, ``metallic'' periodic boundary conditions are applied~\cite{cerda2008p}. Each particle $i$ is characterized by the position vector
\begin{eqnarray}\textbf{r}_i = r_i \hat{\textbf{r}}_i =r_i(\sin \theta_i \cos \phi_i, \sin \theta_i \sin \phi_i, \cos \theta_i)
\end{eqnarray}
\noindent and magnetic moment
\begin{eqnarray}\label{m_omega}
\bm{m}_i =m_i \bm{\Omega}_i =m_i (\sin \omega_{i} \cos \xi_{i}, \sin \omega_{i} \sin \xi_{i}, \cos\omega_{i}).
\end{eqnarray}
\noindent The magnetic interaction between particles \emph{i} and \emph{j} is defined by the anisotropic dipole-dipole interaction $U_d$
\begin{eqnarray}\label{U_d}
U_{d}(ij) = \frac{ (\bm{m}_{i} \cdot \bm{m}_{j})}{r_{ij}^{3}} - 3 \frac{(\bm{m}_{i} \cdot \textbf{r}_{ij}) (\bm{m}_{j} \cdot \textbf{r}_{ij})}{r_{ij}^{5}},
\end{eqnarray}
\noindent where $\textbf{r}_{ij} = \textbf{r}_{j} - \textbf{r}_{i}$ is the interparticle separation vector and $r_{ij} = | \textbf{r}_{ij}|$. Since the distance between particles \emph{i} and \emph{j} can be no smaller than the lattice period $a\geqslant \sigma$, it is convenient to introduce the effective dipolar coupling constant $\lambda_e$ for cubic lattice as follows
\begin{eqnarray}\label{Lambda_e}
\lambda_e = \frac{m^2 \beta}{a^3},
\end{eqnarray}
\noindent which measures the importance of the magnetic interaction of two particle as compared to the thermal energy.

In a uniform external magnetic field \emph{H}, the total interaction energy can be written in units of the thermal energy as
\begin{eqnarray}\label{U}
\beta U = \sum _ {i=1} ^ {N-1} \sum _ {j>i} ^ N \beta U_{d}(ij) - \sum_{i=1}^N \alpha \cos \omega_i,
\end{eqnarray}
\noindent where the Langevin parameter $\alpha = \beta m H$ shows the relation of the Zeeman interaction to the thermal energy.

\subsection{Theory}\label{Theory}

\subsubsection{The Helmholtz free energy expansion of the SCLF in an applied field}
The Helmholtz free energy $F$ of the SCLF at external magnetic field can be presented as the sum of the Helmholtz free energy of an ideal paramagnetic system of non-interacting particles $F_{\textrm{id}}$ and the configurational part $\Delta F$, which takes into account dipole-dipole interparticle interactions
\begin{eqnarray}\label{F}
\beta F = \beta  F_{id} + \beta  \Delta F = - N \ln \left( \frac{\sinh \alpha } { \alpha}\right) + \beta \Delta F.
\end{eqnarray}
\noindent The last term $\Delta F$ is defined via the ratio of the configurational integral of the SCLF, including dipole-dipole interacrion \emph{Z} and the configurational integral of an ideal paramagnetic system of non-interacting particles $Z_{\textrm{id}}$
\begin{eqnarray}\label{dF}
\beta  \Delta F &=& - \ln \left( \frac{Z } {Z_{\textrm{id}}}\right),\\
\frac{Z } {Z_{\textrm{id}}}&=& \frac { \prod \limits _  {k=1}^N \int p(\textbf{r}_k) d \textbf{r}_k d \bm{\Omega}_k }{\prod \limits _  {i=1}^N\int p(\textbf{r}_i) d \textbf{r}_i d \bm{\Omega}_i \exp \left( \sum \limits _{j=1}^N \alpha \cos \omega_j \right)}\nonumber\\
&&\times\exp \left( -\beta \sum \limits _  {i=1} ^ {N-1} \sum \limits _  {i<j} ^ N U_d(ij) + \sum \limits_{j=1}^N \alpha \cos \omega_j \right),\label{Z} \\
d \textbf{r}_i &=& r_i^2  \sin \theta_i d r_i d \theta_i d\phi_i ,\\
d \bm{\Omega}_i &=& \frac{1}{4 \pi}\sin \omega_i d\omega_i d\xi_i,
\end{eqnarray}
\noindent where $p(\textbf{r}_i)$ is the probability distribution function (PDF) of the position of particle \emph{i}. It should be noted that, with this definition, it does not matter if the model system is a fluid or a solid. For a fluid, the PDF is
\begin{eqnarray}\label{PDF_fluid}
p(\textbf{r}_i) = \frac{1}{V},
\end{eqnarray}
\noindent while for a solid
\begin{eqnarray}\label{PDF_solid}
p(\textbf{r}_i) = \delta (\textbf{r}_i - \textbf{r}_i^{(0)}),
\end{eqnarray}
\noindent where $\textbf{r}_i^{(0)}$ is the 'lattice position' of particle \emph{i}. This lattice position can be in a crystalline lattice, or in a random configuration. In all cases, the normalization condition is as follows
\begin{eqnarray}\label{PDF_norm}
\int p(\textbf{r}_i) d \textbf{r}_i= 1.
\end{eqnarray}
In the denominator of Eq. (\ref{Z}), the integrand function depends only on the orientation of magnetic moments. This allows us to obtain the result of integration
\begin{eqnarray}\label{Zid}
\prod \limits _{i=1} ^ N \int p(\textbf{r}_i) d \textbf{r}_i d \bm{\Omega}_i \exp \left( \alpha \cos \omega_i \right) = \left(\frac{\sinh \alpha}{\alpha}\right)^N.
\end{eqnarray}
\noindent Using the Boltzmann-weighted integration over the orientation of particle \emph{i}
\begin{eqnarray}\label{Psi}
d \bm{\Psi}_i = \left(\frac{ \alpha}{\sinh\alpha}\right) \exp \left( \alpha \cos \omega_i \right) d \bm{\Omega}_i,
\end{eqnarray}
\noindent it is possible to reduce definition (\ref{Z}) as
\begin{eqnarray}\label{Znew}
\frac{Z } {Z_{\textrm{id}}}&=& \prod \limits _  {k=1} ^ N\int p(\textbf{r}_k) d \textbf{r}_k d \bm{\Psi}_k  \prod \limits _  {i<j} (1+f_{ij}), \\
f_{ij} &=& \exp \left(-\beta U_d(ij)\right) - 1,
\end{eqnarray}
\noindent where $f_{ij}$ is the Mayer function. In this article, we will take into account only interparticle interactions in all ferroparticle pairs that corresponds with the second virial coefficient level. This means that the product of Mayer functions in Eq. (\ref{Znew}) should be expanded as the following sum
\begin{eqnarray}\label{Zb2}
\frac{Z } {Z_{\textrm{id}}}&=& \prod \limits _  {k=1} ^ N \int p(\textbf{r}_k) d \textbf{r}_k d \bm{\Psi}_k \times \left(1+  \sum \limits _  {i<j}f_{ij}\right) \nonumber \\
&=& 1 + \sum \limits _  {i<j} \int p(\textbf{r}_i) d \textbf{r}_i d \bm{\Psi}_i \int p(\textbf{r}_j) d \textbf{r}_j d \bm{\Psi}_j f_{ij}.
\end{eqnarray}
\noindent Using the definition of PDF (\ref{PDF_solid}), one can obtain for the SCLF
\begin{eqnarray}\label{Zb2_vir}
\frac{Z } {Z_{\textrm{id}}}&=& 1 + \sum \limits _  {i<j} \int d \bm{\Psi}_i d \bm{\Psi}_j f_{ij}^{(0)} = 1 + \sum \limits _  {i<j} \left\langle f_{ij}^{(0)}\right\rangle,
\end{eqnarray}
\noindent where $f_{ij}^{(0)} = \int p(\textbf{r}_i) d \textbf{r}_i \int p(\textbf{r}_j)d \textbf{r}_j f_{ij} $ is the Mayer function in the 'lattice position' of particles \emph{i} and \emph{j} and the angle brackets $\left\langle \ldots\right\rangle$ mean a Boltzmann-weighted average (\ref{Psi}) over the orientation of each particles involved. The configurational part of the Helmholtz free energy in approximation by the linear term of the logarithm  expansion looks like
\begin{eqnarray}\label{df_sum}
\frac{\beta \Delta F} {N} = - \frac{1}{N} \ln \left ( 1 + \sum \limits _  {i<j} \left\langle f_{ij}^{(0)}\right\rangle \right)&\simeq& - \frac{1}{N}\sum \limits _  {i<j} \left\langle f_{ij}^{(0)}\right\rangle.
\end{eqnarray}
\noindent The right-hand side of Eq. (\ref{df_sum}) needs some discussion: for each value of \emph{i}, the sum $\sum _{i<j} \left\langle f_{ij}^{(0)}\right\rangle$ will be finite; and dividing the sum over  \emph{i} by \emph{N} gives a number that does not depend on \emph{N}:
\begin{eqnarray}\label{sum}
\frac{1}{N} \sum \limits _  {i<j} \left\langle f_{ij}^{(0)}\right\rangle = \frac{1}{2N}\sum \limits _  {i=1} ^N  \sum \limits _  {j\neq i}^N \left\langle f_{ij}^{(0)}\right\rangle = \frac{1}{2}\sum \limits _  {j = 2} ^N \left\langle f_{1j}^{(0)}\right\rangle.
\end{eqnarray}
\noindent The final results for $\Delta F$ of the SCLF is
\begin{eqnarray}\label{dF_vir}
\frac{\beta \Delta F} {N}&=& - \frac{1}{2}\sum \limits _  {j=2} ^N \left\langle f_{1j}^{(0)}\right\rangle.
\end{eqnarray}
\noindent The absence of the numerical concentration $\rho$ in Eq. (\ref{dF_vir}) means that the properties of the SCLF with a regular cubic lattice scale in a simple way with the volume.

For a ferrofluid, which is modeled by the system of moving dipolar hard spheres (DHS), it is possible to write the configurational part of the Helmholtz free energy on the second virial coefficient level in the following form \cite{Balescu1975}
\begin{eqnarray}
\frac{\beta \Delta F^{\textrm{\scriptsize{DHS}}}} {N}&=& \rho B_2^{\textrm{\scriptsize{DHS}}},\label{dF_b2_fluid}\\
B_2^{\textrm{\scriptsize{DHS}}} &=& - \frac{1}{2} \int d \textbf{r}_{12}  \left\langle f_{12}\right\rangle,\label{b2_fluid}
\end{eqnarray}
\noindent where hard-sphere condition $r_{12}\geqslant \sigma$ is assumed.  Representation of the integral in Eq. (\ref{b2_fluid}) via the integral sum over cubic lattice nodes with $\Delta V = a^3$ and substitution of $\rho=a^{-3}$ into Eq. (\ref{dF_b2_fluid}) give a result coinciding with (\ref{dF_vir}).

\subsubsection{Discrete position averaging for the SCLF}

In this article, we will consider the expansion of the Mayer function into a series over the powers of dipolar energy up to $U_d^3$
\begin{eqnarray}
f_{1j}^{(0)} = -\beta U_d (1j) + \frac{1}{2!} \left[ -\beta U_d(1j) \right]^2 + \frac{1}{3!} \left[ -\beta U_d (1j) \right]^3.
\end{eqnarray}
\noindent Dipole-dipole potential $U_d$ depends on the radius vector $\textbf{r}_{1j}$ of particles 1 and \emph{j} and the magnetic moments $\textbf{m}_{1}$ and $\textbf{m}_{j}$. After the Boltzmann-weighted averaging over the orientation of magnetic moments $\textbf{m}_{1}$ and $\textbf{m}_{j}$, one can write $\Delta F$ in the following way

\begin{eqnarray}\label{b2Lambda3}
\frac{\beta \Delta F} {N} &=& - \frac{1}{2} \left(b_1 \lambda_e + b_2 \lambda_e^2 + b_3 \lambda_e^3 \right),\\
b_1 &=&\sum \limits _  {j=2} ^N \bigg\langle \frac{-\beta U_d (1j)}{\lambda_e}\bigg\rangle =  2 L^2(\alpha) \gamma_{12}, \nonumber \\
b_2 &=& \sum \limits _  {j=2} ^N \bigg\langle \frac{1}{2!} \left(\frac{ -\beta U_d(1j)}{\lambda_e} \right)^2\bigg\rangle =  \frac{36}{35} L_3^2(\alpha) \gamma_{24} \nonumber \\
&+& \frac{2}{3} L_3(\alpha) \left( 1-\frac{L_3(\alpha)}{7} \right) \gamma_{22} + \frac{1}{3} \left(1+ \frac{L_3^2(\alpha)}{5} \right) \gamma_{20}, \nonumber \\
b_3 &=&  \sum \limits _  {j=2} ^N \bigg\langle \frac{1}{3!} \left(\frac{ -\beta U_d(1j)}{\lambda_e} \right)^3\bigg\rangle\nonumber \\
&=&\frac{24}{77} \left( L^2(\alpha) - 10 \frac{L(\alpha)L_3(\alpha)}{\alpha} + 25 \frac{L^2_3(\alpha)}{\alpha^2}\right)\gamma_{36}  \nonumber \\
 &+& \frac{72}{385} \left( 2 L^2(\alpha) - 9 \frac{L(\alpha)L_3(\alpha)}{\alpha} - 5 \frac{L^2_3(\alpha)}{\alpha^2}\right) \gamma_{34} \nonumber \\
 &+& \frac{4}{7} \left( L^2(\alpha) - \frac{L(\alpha)L_3(\alpha)}{\alpha} + \frac{L^2_3(\alpha)}{\alpha^2}\right) \gamma_{32}  \nonumber \\
 &+& \frac{2}{105} \left( 4 L^2(\alpha) + 2 \frac{L(\alpha)L_3(\alpha)}{\alpha} - 5 \frac{L^2_3(\alpha)}{\alpha^2}\right) \gamma_{30} , \nonumber\\
\gamma_{pq} &=& \sum \limits _  {j=2} ^N \frac{1}{\tilde{r}_{1j}^{3p}} P_q \left( \frac{\tilde{z}_{1j}}{\tilde{r}_{1j}} \right),\nonumber
 \end{eqnarray} 
 
%\gamma_{12} &=& \sum \limits _  {j=2} ^N \frac{1}{\tilde{r}_{1j}^3} P_2 \left( \frac{\tilde{z}_{1j}}{\tilde{r}_{1j}} \right), \ \ \ \ \gamma_{36} = \sum \limits _  {j=2} ^N \frac{1}{\tilde{r}_{1j}^9} P_6 \left( \frac{\tilde{z}_{1j}}{\tilde{r}_{1j}}  \right), \nonumber\\
%\gamma_{24} &=& \sum \limits _  {j=2} ^N \frac{1}{\tilde{r}_{1j}^6} P_4 \left( \frac{\tilde{z}_{1j}}{\tilde{r}_{1j}}  \right), \ \ \ \ \gamma_{34} = \sum \limits _  {j=2} ^N \frac{1}{\tilde{r}_{1j}^9} P_4 \left( \frac{\tilde{z}_{1j}}{\tilde{r}_{1j}} \right) ,\nonumber\\
%\gamma_{22} &=& \sum \limits _  {j=2} ^N  \frac{1}{\tilde{r}_{1j}^6}   P_2 \left( \frac{\tilde{z}_{1j}}{\tilde{r}_{1j}} \right) , \ \ \ \ \gamma_{32} = \sum \limits _  {j=2} ^N \frac{1}{\tilde{r}_{1j}^9} P_2 \left( \frac{\tilde{z}_{1j}}{\tilde{r}_{1j}} \right), \nonumber\\
%\gamma_{20} &=&\sum \limits _  {j=2} ^N \frac{1}{\tilde{r}_{1j}^6} P_0 \left( \frac{\tilde{z}_{1j}}{\tilde{r}_{1j}} \right) ,\ \ \ \ \gamma_{30} = \sum \limits _  {j=2} ^N \frac{1}{\tilde{r}_{1j}^9} P_0 \left( \frac{\tilde{z}_{1j}}{\tilde{r}_{1j}} \right),\nonumber\\
%L(\alpha) &=& \coth \alpha - \frac{1}{ \alpha},  \ \ \ \ \ \ \  L_3(\alpha) = 1- 3 \frac{L(\alpha) }{ \alpha},\nonumber

\noindent where $ L(\alpha)$ is the Langevin function, $P_q$ $(q=0,2,4,6)$ are the Legendre polynomials, $\tilde{\textbf{r}}_{1j} = \textbf{r}_{1j} / a $  is a dimensionless center-center vector, $\tilde{z}_{1j}$ is \emph{z}-component of vector $\tilde{\textbf{r}}_{1j}$ in a coordinate system, where \emph{Oz} axis is selected parallel to the direction of the external magnetic field \textrm{\textbf{H}}. The details of the Boltzmann-weighted averaging of the powers of dipolar energy over the orientation of magnetic moments $\textbf{m}_{1}$ and $\textbf{m}_{j}$ can be found in supplemental materials of Ref. \cite{Elfimova2013}
\begin{table*}[t!]
\begin{center}
\caption{\label{tabl_solid} The results of the calculation of numbers $\gamma_{pq}$ for the SCLF. \emph{R} is the dimensionless radius of the cylinder; \emph{h} is the factor describing the cylinder elongation so that the total cylinder height is $H = 2hR$; \emph{N} is the number of ferroparticles.  }
\smallskip
\begin{tabular}{@{} c | *{7}{c|} c}
\hline
\hline
& \multicolumn{2}{c|}{$R=1$} & \multicolumn{2}{c|}{$R=2$} & \multicolumn{2}{c|}{$R=10$} & \multicolumn{2}{c}{$R=20$}\\
\cline{2-9}
\raisebox{0.5ex}[0cm][0cm]{$\gamma_{pq}$} & $h=10$  & $h=500$ & $h=10$  & $h=500$  & $h=10$ & $h=500$ & $h=10$ & $h=500$  \\
 & $ \ \ N=105\ \ $  & $N=5005$ & \ \ $N=533$ \ \ & $N=26013$  & $N=63717$ & $N=3170317$ & $N= 504057$ & $N=25141257$  \\
\hline
$\gamma_{12}$ & 2.0683 & 2.1131 & 2.0639 & 2.0946 & 2.0632 & 2.0944 & 2.0634 & 2.0944\\
$\gamma_{24}$ &3.1458 & 3.1458 & 3.2182 & 3.2182& 3.2257 & 3.2257 & 3.2257 & 3.2257\\
$\gamma_{22}$ & 0.3385 & 0.3385& 0.0761 & 0.0761 & 0.0006 & 0.0006 & 0.0001 & 0.0001 \\
$\gamma_{20}$ & 7.1090 & 7.1091& 8.0978 & 8.0978 & 8.3995 & 8.3995 & 8.4016 &8.4016\\
$\gamma_{36}$ & 0.7000 & 0.7000 & 0.6557 & 0.6557 & 0.6553 & 0.6553 & 0.6553 & 0.6553\\
$\gamma_{34}$ & 3.3616 &  3.3616 & 3.4068 & 3.4068 & 3.4081 & 3.4081 &3.4081 & 3.4081\\
$\gamma_{32}$ & 0.0967 & 0.0967 & 0.0048 & 0.0048 & 0.0000 & 0.0000 & 0.0000 & 0.0000\\
$\gamma_{30}$ & 6.3636 & 6.3636 & 6.6144 & 6.6144 & 6.6289 & 6.6289 & 6.6289 & 6.6289\\
\hline
\hline
\end{tabular}
\end{center}
\end{table*}
To calculate the numbers $\gamma_{pq}$, it is necessary to perform summation by substituting the specific coordinates of particles \emph{1} and \emph{j}. Let us fix particle \emph{1} at the origin of the laboratory coordinate system, meaning $(\tilde{x}_{1}, \tilde{y}_{1},\tilde{z}_{1}) \equiv (0,0,0)$. The coordinates of particle \emph{j} in this case can be located in all the other nodes of the cubic lattice, limited by cylinder size:
\begin{eqnarray}
-R \leqslant & \tilde{x}_{j}&\leqslant R,\nonumber  \\
-R \leqslant& \tilde{y}_{j}&\leqslant R, \nonumber \\
-h R \leqslant& \tilde{z}_{j}&\leqslant h R,\nonumber\\
\left(\tilde{x}_{j}\right)^2 + & \left(\tilde{y}_{j}\right)^2 & \leqslant R^2,\nonumber\\
\left(\tilde{x}_{j}\right)^2 + & \left(\tilde{y}_{j}\right)^2&+\left(\tilde{z}_{j}\right)^2>0.\nonumber
\end{eqnarray}
\noindent where \emph{R} is the dimensionless radius of the cylinder and \emph{h} is the factor describing the cylinder elongation. First, let us to show the convergence of the series in numbers $\gamma_{pq}$. Each $\gamma_{pq}$ contains the Legendre polynomial $P_q$, the argument of which is limited as
\begin{eqnarray}
\left |  \frac{\tilde{z}_{1j}}{\tilde{r}_{1j}} \right | \leqslant 1.
\end{eqnarray}
\noindent Thus, each Legendre polynomial obeys the following relation
\begin{eqnarray}
\left |P_q \left( \frac{\tilde{z}_{1j}}{\tilde{r}_{1j}} \right)\right | \leqslant 1
\end{eqnarray}
\noindent and numbers $\gamma_{pq}$ can be estimated as
\begin{eqnarray}
\left |\gamma_{pq}  \right | \leqslant  \sum \limits _  {j=2} ^N \frac{1}{\tilde{r}_{1j}^{3p}} \leqslant  \sum \limits _  {j=2} ^N \frac{1}{|\tilde{z}_{j}|^{3p}} \leqslant 2 R^2 \sum \limits _  {n=1} ^{hR} \frac{1}{n^{3p}}.
\end{eqnarray}
\noindent The thermodynamic limit in our model means that at the constant radius of cylinder $R$, its elongation $h \rightarrow \infty$; therefore, in this case one can obtain
\begin{eqnarray}
\left |\gamma_{pq}  \right |  \leqslant 2 R^2 \sum \limits _  {n=1} ^{\infty} \frac{1}{n^{3p}} = 2 R^2 \zeta (3p),
\end{eqnarray}
\noindent where $\zeta(s)$-function converges absolutely and evenly for
$s \geqslant 1 + \varepsilon$, $\varepsilon>0$.

Several different systems were considered to show how the numbers $\gamma_{pq}$ depend on the cylinder size and form.
The results are summarized in Table \ref{tabl_solid}.
One can note some dependence of $\gamma_{12}$ on the cylinder elongation,
while the other numbers $\gamma_{pq}$ demonstrate the same values for $h=10$ and $h=500$.
This means that the long-range contribution $\gamma_{12} \sim \tilde{r}_{1j}^{-3}$ is sensitive to demagnetization effects due to the lowest rate of decrease with distance, unlike other values $\gamma_{pq}$.
A similar fact has already been discussed in \citep{Elfimova2013}, which is devoted to the thermodynamics of ferrofluids in applied magnetic fields.
Data from Table \ref{tabl_solid} show that the averaging of the second Legendre polynomial in numbers $\gamma_{22}$ and $\gamma_{32}$ gives zero results. Since values of $\gamma_{pq}$ for the system with $R=10$ and $R=20$ are very close, it is assumed that the biggest system, with $N = 25141257$ particles, is enough for numerical calculation of $\gamma_{pq}$.

The final analytical expression for the $\Delta F$ expansion can be presented as
\begin{eqnarray}
\frac{\beta \Delta F} {N}=  - \frac{\pi}{6} \Bigg[ 4 \lambda_e L^2(\alpha) +  \lambda_e^2 \left( 2.674+ 3.703 L_3^2(\alpha) \right)&&\nonumber \\
+ \lambda_e^3 \left( 1.895 L^2(\alpha) -  \frac{ 7.187 L(\alpha)L_3(\alpha)}{\alpha} +  \frac{1.230 L^2_3(\alpha)}{\alpha^2}\right)  \Bigg] . && \label{F_solid_final}
\end{eqnarray}

\noindent Formula (\ref{F_solid_final}) will be referred to as the Virial free energy (VFE) theory. To avoid the strong dependence of the Helmholtz free energy on the effects of truncation at the low order in  $\lambda_e$-expansion, it is useful to apply the logarithmic transformation of  $\Delta F$ according to the initial definition (\ref{dF}):
\begin{eqnarray}
\frac{\beta \Delta F } {N}&=& - \ln \Bigg\{ 1 + \frac{\pi}{6} \bigg[ 4 \lambda_e L^2(\alpha)\nonumber \\
&+&  \lambda_e^2 \left( 2.674+ 3.703 L_3^2(\alpha) \right)\nonumber\\
 &+& \lambda_e^3 \bigg( 1.895 L^2(\alpha) -  \frac{ 7.187 L(\alpha)L_3(\alpha)}{\alpha} \nonumber \\
 &+&  \frac{1.230 L^2_3(\alpha)}{\alpha^2}\bigg)  \bigg] \Bigg\} .\label{F_solid_log}
 \end{eqnarray}
\noindent The obtained expression (\ref{F_solid_log}) will be referred to as the Logarithmic free energy (LFE) theory.  In Refs. \cite{Elfimova2012,Elfimova2013,Solovyova2020}, it was shown that this method of the Helmholtz free energy logarithmic transformation is capable of considerably expanding the theory's applicability range over concentration and intensity of interparticle dipole-dipole interactions for describing thermodynamic and magnetic properties of dipolar hard spheres fluids.

\begin{figure*}[t]
\center
\includegraphics[width=0.8\linewidth]{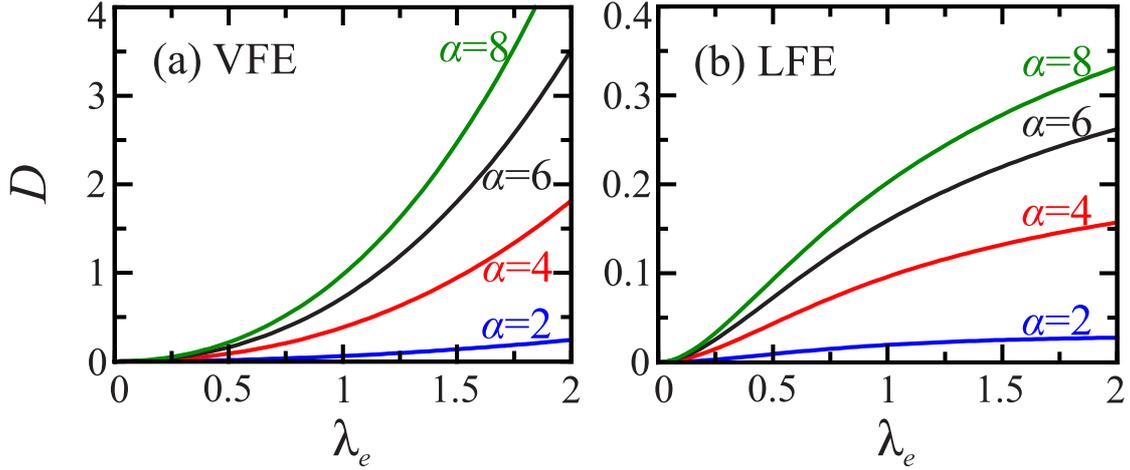}
\caption{The value of $D$ as a function of dipolar coupling constant $\lambda_e$ using (a) VFE theory and (b) LFE theory. Lines correspond to $\alpha = 2$, $\alpha = 4$, $\alpha = 6$ and $\alpha = 8$ from bottom to top.}
\label{fig:D}
\end{figure*}
\subsubsection{Comparing the SCLF and the DHS theories}

In this section, the analytical results for both the SCLF and the DHS models are compared with each other. Fluid of DHSs is described by two basic parameters: the numerical concentration of ferroparticles $\rho$ and the dipolar coupling constant
\begin{eqnarray}\label{Lambda}
\lambda = \frac{m^2 \beta}{\sigma^3}.
\end{eqnarray}
\noindent The virial expansion of the Helmholtz free energy expansion for the DHS model up to the second virial coefficient in an external magnetic field has the following form \cite{Elfimova2013}:
\begin{eqnarray}
\frac{\beta } {N}\Delta F^{\textrm{\scriptsize{DHS}}}=  - \frac{\rho \pi \sigma^3}{6} \bigg[ 4 \lambda L^2(\alpha) +  \frac{4\lambda^2}{3} \left(1+ \frac{L_3^2(\alpha)}{5} \right) \nonumber&& \\
 +\frac{4\lambda^3}{105} \left( 4 L^2(\alpha) +  \frac{2L(\alpha)L_3(\alpha)}{\alpha} -  \frac{5L^2_3(\alpha)}{\alpha^2}\right)  \bigg].&&\label{Fold_fluid_vir}
\end{eqnarray}
\noindent If we set the same numerical concentration $\rho =a^{-3}$ as with SCLF, then the configurational part of the Helmholtz free energy of a ferrofluid in notation $\lambda_e$ (\ref{Lambda_e}) can be presented as:
\begin{eqnarray}
\frac{\beta } {N}\Delta F^{\textrm{\scriptsize{DHS}}}=  - \frac{\pi}{6} \bigg[ 4 \lambda_e L^2(\alpha) +  \frac{4\lambda_e^2}{3} \left( \frac{a^3}{\sigma^3} \right) \left(1+ \frac{L_3^2(\alpha)}{5} \right) && \nonumber \\
 + \frac{4\lambda_e^3}{105} \left( \frac{a^3}{\sigma^3} \right)^2 \left( 4 L^2(\alpha) +  \frac{2L(\alpha)L_3(\alpha)}{\alpha} -  \frac{5L^2_3(\alpha)}{\alpha^2}\right)  \bigg].&& \label{F_fluid_vir}
 \end{eqnarray}
\noindent This formula depends on two parameters: the effective dipolar coupling constant defined especially for the SCLF model in (\ref{Lambda_e}), and the relation $a^3/\sigma^3$. The latter occurs due to the transition from the usual definition of the dipolar coupling constant $\lambda$ (\ref{Lambda}) in the DHS model to the new notation of $\lambda_e$ (\ref{Lambda_e}). When using the SCLF theory (\ref{F_solid_final}), it is not necessary to know the lattice period $a$ or particle diameter $\sigma$ separately. Hence for each value of $\lambda_e$, there is no unique way to define the value of $a^3/\sigma^3$, which is necessary for comparing the two models. To determine the value $a^3/\sigma^3$, it is convenient to study the weak field behavior of $\Delta F$ for the SCLF and the DHS models
\begin{eqnarray}
\frac{\beta } {N}\Delta F^{\textrm{\scriptsize{DHS}}} (\alpha \rightarrow 0)&=&  - \frac{\pi}{6} \frac{4}{3} \left( \frac{a^3}{\sigma^3} \right) \lambda_e^2 \nonumber \\
&=& -  0.222\pi \lambda_e^2 \left( \frac{a^3}{\sigma^3} \right),\label{F_fluid_zero}\\
\frac{\beta } {N}\Delta F^{\textrm{\scriptsize{SCLF}}} (\alpha \rightarrow 0)&=&  - \frac{\pi}{6} 2.674 \lambda_e^2 = - 0.446 \pi \lambda_e^2 ,\label{F_solid_zero}
\end{eqnarray}
\noindent where $\Delta F^{\textrm{\scriptsize{SCLF}}}$ is from VFE theory (\ref{F_solid_final}) and $\Delta F^{\textrm{\scriptsize{DHS}}}$ is from  (\ref{F_fluid_vir}). To obtain the equality of $\Delta F^{\textrm{\scriptsize{SCLF}}} \simeq \Delta F^{\textrm{\scriptsize{DHS}}}$ at $\alpha \simeq 0$, one can define the relation $a^3/\sigma^3 \simeq 2 $: this means  $a \simeq 1.26 \sigma$. Volume concentration $\varphi = \pi \rho \sigma^3 / 6$ in this system is equal to $\varphi = 0.262$, that corresponds to very dense DHS fluid. Next, the lattice period will be fixed as $a \simeq 1.26 \sigma$, and we will change either the Langevin parameter $\alpha$ or thermal energy included in $\lambda_e$. In this case, the asymptotes for an infinitely strong field are
\begin{eqnarray}
\frac{\beta } {N}\Delta F^{\textrm{\scriptsize{DHS}}} (\alpha \rightarrow \infty)&=&  - \frac{\pi}{6} \left( 4\lambda_e + 1.6 \lambda_e^2 + 0.61 \lambda_e^3 \right) ,\label{F_fluid_inf}\\
\frac{\beta } {N}\Delta F^{\textrm{\scriptsize{SCLF}}} (\alpha \rightarrow \infty)&=&  - \frac{\pi}{6} \big( 4\lambda_e + 6.378 \lambda_e^2 \nonumber \\
&+& 1.895 \lambda_e^3 \big) .\label{F_solid_inf}
\end{eqnarray}

\noindent This shows that coefficients in $\lambda_e$-expansion of $\Delta F^{\textrm{\scriptsize{DHS}}}$ decrease faster than those of  $\Delta F^{\textrm{\scriptsize{SCLF}}}$. One can conclude from the analytical results that the DHS system is less sensitive to cooling inside a system under strong external magnetic field, while the SCLF is more responsive to the same.

As was shown in \cite{Elfimova2013}, the logarithmic form of free energy
\begin{eqnarray}
\frac{\beta } {N}\Delta F^{\textrm{\scriptsize{DHS}}} = - \ln \Bigg\{ 1  + \frac{\pi}{6} \bigg[ 4 \lambda_e L^2(\alpha) +  \frac{8\lambda_e^2}{3} \left(1+ \frac{L_3^2(\alpha)}{5} \right) &&\nonumber \\
 +\frac{16\lambda_e^3}{105} \left( 4 L^2(\alpha) +  \frac{2L(\alpha)L_3(\alpha)}{\alpha} -  \frac{5L^2_3(\alpha)}{\alpha^2}\right)  \bigg] \Bigg\} &&\label{F_fluid_log}
\end{eqnarray}
is better for describing the thermodynamic properties of ferrofluids. In Eq. (\ref{F_fluid_log}), the notations of this paper are used and $a \simeq 1.26 \sigma$ is assumed. The relative difference of $\Delta F$ for two models
\begin{eqnarray}
D &=& \bigg | \frac{\beta } {N}\Delta F^{\textrm{\scriptsize{SCLF}}} - \frac{\beta } {N}\Delta F^{\textrm{\scriptsize{DHS}}} \bigg | \label{F_dif}
 \end{eqnarray}
as function of $\lambda_e$ is presented in Fig. \ref{fig:D}. Fig. \ref{fig:D} (a) shows the relative difference of virial expansions of free energy for the SCLF (\ref{F_solid_final}) and ferrofluids (\ref{F_fluid_vir}) at the various values of the Langevin parameter. Note that $D \equiv 0 $ at $\alpha = 0$ due to the assumption $a \simeq 1.26 \sigma$. As $\alpha$ increases, the relative difference \emph{D} goes up. In a weak field $\alpha < 2$, this difference is not significantly pronounced, but in a medium field $2<\alpha < 6$, \emph{D} rapidly increases. With a further increase of $\alpha$, the value of \emph{D} grows more slowly. Fig. \ref{fig:D} (b) shows the relative difference of the logarithmic form of free energy for the SCLF (\ref{F_solid_log}) and ferrofluids (\ref{F_fluid_log}). It should be noted that the scale on the second graph is ten times larger than the same on the first graph. This is because the virial series on Fig. \ref{fig:D} (a) is more sensitive to changing of coefficients in terms. It is seen that at $\alpha = 2$ and $\lambda_e \sim 1.5$, the relative difference \emph{D} reaches a plateau: it no longer changes its value. For  $\alpha > 2$, the value of $D$ slows the rate of increase at large values of $\lambda_e>2$.

\begin{figure}[t!]
\center
\includegraphics[width=\linewidth]{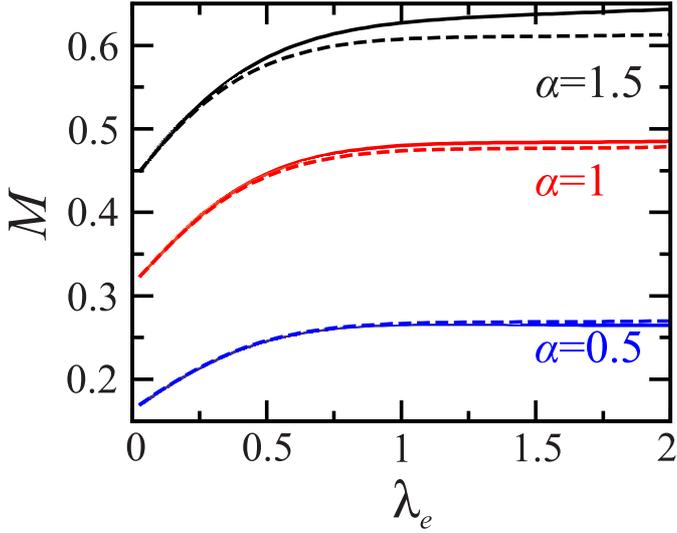}
\caption{The scalar magnetization $M$ as a function of dipolar coupling constant $\lambda_e$ using LFE theory for SCLF (\ref{F_solid_log}) (solid lines) and DHS (\ref{F_fluid_log}) (dashed lines). Lines correspond to $\alpha = 0.5$, $\alpha = 1$ and $\alpha = 1.5$ from bottom to top.}
\label{fig:mag_c}
\end{figure}

Now let us compare the magnetization of the SCLF and the DHS systems. In theory, the fractional (scalar) magnetization can be defined via the Helmholtz free energy as
\begin{eqnarray}\label{mag}
 M &=& - \frac{\partial}{\partial \alpha} \left( \frac{\beta  F}{N} \right) = M_L - \frac{\partial}{\partial \alpha} \left( \frac{\beta \Delta F}{N} \right).
\end{eqnarray}
\noindent In Fig. \ref{fig:mag_c}, the magnetization curves for both the SCLF and DHS models are given from the LFE theory. It should be noted that the DHS magnetization curve is higher than that of the SCLF system in a weak field $\alpha \lesssim 0.5$, and lower in moderate and strong magnetic fields. Additional increases to the magnetization of the DHS fluid in a weak magnetic field can be explained by the formation of dimers or chains aligned in the magnetic field's direction due to the dipole-dipole interaction \cite{Elfimova2012_JCP}, while the appearance of dimers in the SCLF system with immobilized particles is impossible. For $\alpha > 0.5$ in the DHS fluid with strong and moderate dipole-dipole interactions, one can observe the formation of correlation structures in which particle magnetic moments can compensate for each other. This leads to lower growth in magnetization compared to the SCLF model.

\subsection{Molecular dynamics simulations}

To test the accuracy and application limits of the developed SCLF theories,
we thoroughly tested their predictions against molecular dynamics (MD) simulations.
The modelled system consists of $N$ identical ferroparticles
rigidly fixed at the lattice nodes inside a cubic simulation cell with
3D periodic boundary conditions imposed.
The rotational dynamics of the $i$-th particle is governed by the Langevin equation
\begin{equation}\label{lang_eq}
\mathcal{I} \frac{d \bm{W}_i}{d t} = - \bm{\Omega}_i \times \frac{\partial U}{\partial \bm{\Omega}_i} - \Gamma \bm{W}_i + \bm{\eta}_i,
\end{equation}

\noindent where $\mathcal{I}$ is the particle moment of inertia,  $\bm{\Omega}_i$ is the unit vector directed along the magnetic moment (\ref{m_omega}),
$\bm{W}_i$ is the particle angular velocity,
i.e., $d \bm{\Omega}_i / dt = \bm{W}_i \times \bm{\Omega}_i$,
$\Gamma$ is the friction coefficient,
$\bm{\eta}_i$ is the thermal noise torque,
which has a zero mean value $\overline{\bm{\eta}_{i}(t)}  = \bm{0}$
and satisfies the fluctuation-dissipation
relationship $\overline{ \eta_{i, k} (t_1)\eta_{j,l}(t_2)} = 2 \beta^{-1} \Gamma \delta_{ij} \delta_{kl} \delta(t_1 - t_2)$,
$k$ and $l$ are Cartesian indices,
$\delta_{ij}$ is the Kronecker delta,
$\delta(t)$ is the Dirac delta function, and
overline means the average over noise realizations.
The simulations were performed using the ESPResSo 4.0 package~\cite{weik2019espresso}.
In~actual simulations, only dimensionless quantities were used,
and $\mathcal{I}$, $\sigma$ and $\beta$ were chosen as units.
The input simulation parameters are
the reduced friction coefficient
$\Gamma^* = \Gamma \sqrt{\beta / \mathcal{I}}$,
the reduced lattice period $a^* = a/\sigma$,
the dipolar coupling constant $\lambda_e$
and the Langevin parameter $\alpha$.
Typically, $\Gamma^* = 1$ and $a^* = 1$,
while $\lambda_e$ and $\alpha$ can vary within broad ranges.
The dimensionless time step is $\Delta t^* = \Delta t/\sqrt{\beta \mathcal{I}} = 0.001$.
The torques due to long-range dipole-dipole interactions are computed using
the dipolar P$^3$M algorithm with ``metallic'' boundary conditions~\cite{cerda2008p}.
Initial orientations of magnetic moments are random.
A typical simulation consisted of $5 \times 10^4$ equilibration time steps,
followed by a production run of at least $2.5 \times 10^5$ time steps.

Assuming that the external magnetic field is directed along the $Oz$ axis,
the fractional magnetization can be calculated in the simulation simply as
\begin{equation}\label{sim_magn}
M = \frac{1}{N}\left\langle \sum_{i = 1}^{N} \cos \omega_i \right\rangle_{t},
\end{equation}
where $\langle \ldots \rangle_t$ means the average over simulation time.
Magnetization values given in Sec.~\ref{sec:rslt} are calculated for $N = 1000$
(i.e., for $10 \times 10 \times 10$ lattice).
Initial magnetic susceptibility can be calculated at $\alpha = 0$
as~\cite{de1986computer}:
\begin{equation}\label{sim_susc}
\chi = \chi_L \Bigg\langle \left(\sum_{i = 1}^{N} \bm{\Omega}_i \right)^2 \Bigg\rangle_{t} \frac{1}{N}.
\end{equation}
Possibility of spontaneous orientational ordering at $\alpha = 0$
was investigated by means of the scalar order parameter
\begin{equation}\label{sim_order}
	S \equiv \left\langle \textrm{Tr} \:\bm{Q}^2 \right\rangle = \frac{3}{2} \left\langle  \frac{1}{N^2} \sum_{i,j} \left(\bm{\Omega}_i \cdot \bm{\Omega}_j\right)^2 - \frac{1}{3}\right\rangle_{t},
\end{equation}
where Tr means trace and $\bm{Q}$ is the orientation tensor, which is well known in the physics of liquid crystals~\cite{de1993physics}:
\begin{equation}\label{orient_tensor}
Q_{kl} = \sqrt{\frac{3}{2}} \left( \frac{1}{N} \sum_{i = 1}^{N} \Omega_{i,k} \Omega_{i,l}  - \frac{1}{3}\delta_{kl}\right),
\end{equation}

\begin{figure*}[t!]
\center
\includegraphics[width=0.8\linewidth]{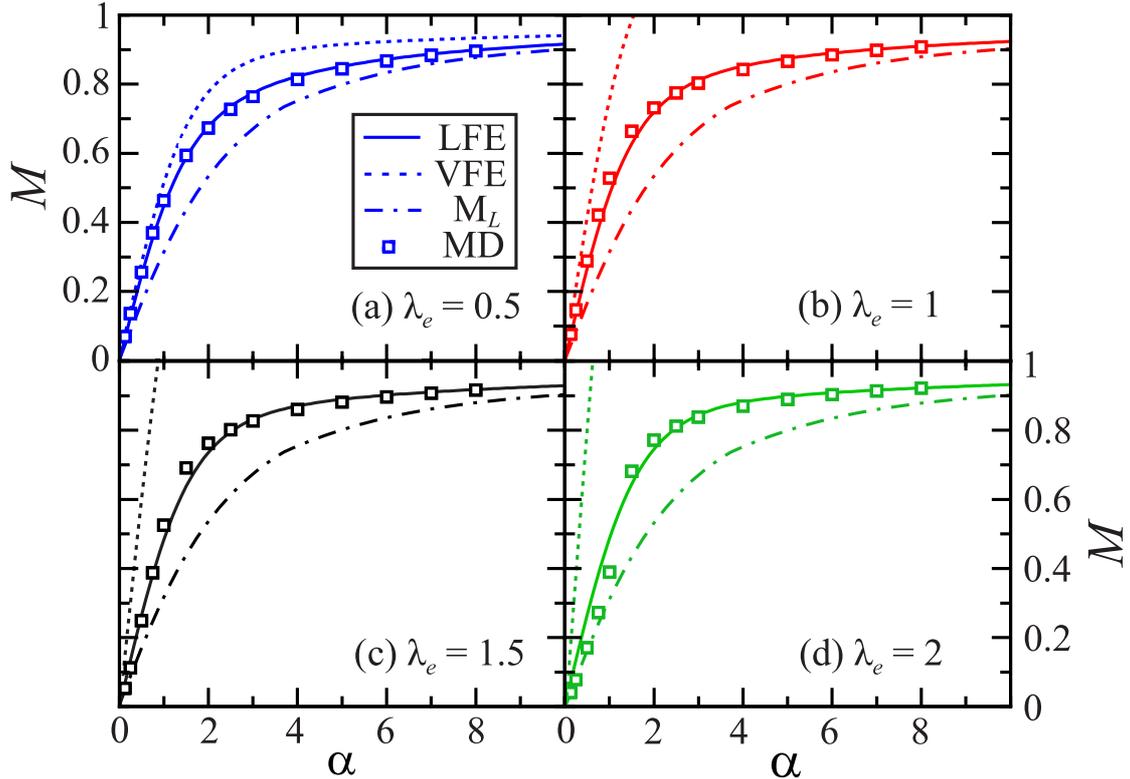}
\caption{The fractional magnetization \emph{M} as a function of Langevin parameter $\alpha$ for a model system with (a) $\lambda_e = 0.5$, (b) $\lambda_e = 1$, (c) $\lambda_e = 1.5$ and (d) $\lambda_e = 2$. Points are from MD simulations, solid lines are from LFE theory (\ref{F_solid_log}), dashed lines are from VFE theory (\ref{F_solid_final}), and dash dotted lines are from the Langevin model [Eq. (\ref{Lang})].}
\label{fig:mag}
\end{figure*}

\noindent where $\Omega_{i,k}$ is the $k$-th component of $\bm{\Omega}_i$.
According to Ref.~\cite{vieillard1974equation}, $S = 1$ corresponds to a completely ordered state,
in which all moments are directed along or against a common direction,
while in the isotropic disoriented state $S$ is of the order of $N^{-1}$.
The numerical results for $\chi$ and $S$ given in Sec.~\ref{sec:rslt}
are obtained for $N = 512$ (i.e., for $8 \times 8 \times 8$ lattice).
Presented values of $\chi$ and $S$ were averaged over four independent simulation runs.

\section{Results}\label{sec:rslt}
\subsection{Comparison between theory and simulation}

\begin{figure*}[t!]
\center
\includegraphics[width=0.8\linewidth]{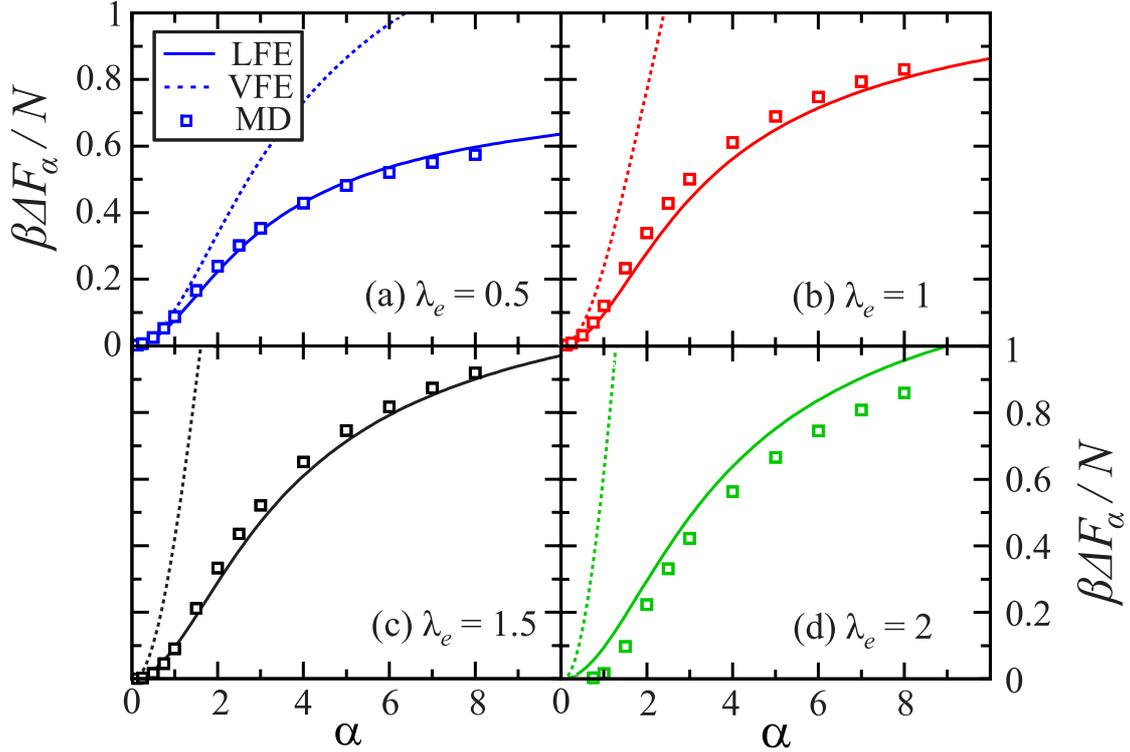}
\caption{The value of $\Delta F _{\alpha} $ as a function of Langevin parameter $\alpha$ for the SCLF with (a) $\lambda_e = 0.5$, (b) $\lambda_e = 1$, (c) $\lambda_e = 1.5$ and (d) $\lambda_e = 2$. Points are from MD simulations, solid lines are from LFE theory (\ref{F_solid_log}), and dashed lines are from VFE theory (\ref{F_solid_final}).}
\label{fig:free_en}
\end{figure*}
Fig. \ref{fig:mag} shows scalar magnetization \emph{M} as a function of Langevin parameter $\alpha$ for model system with $\lambda_e = 0.5$, 1, 1.5 and 2, from MD simulations and theories. The Langevin magnetization $M_L$ (\ref{Lang}) does not take into account the interparticle dipole-dipole interaction, since it can be obtained from Eq. (\ref{mag}) by setting $\Delta F \equiv 0$. However, the MD simulation results demonstrate clearly that ignoring interparticle interactions leads to a great underestimation of magnetization even for the system with the low dipolar coupling constant.
On the other hand, the VFE theory greatly overestimates the MD results, because only pair correlations were taking into account in the virial expansion of the Helmholtz free energy. As was shown earlier, at zero magnetic field a second virial coefficient is positive: this leads to increases in the initial slope of the magnetization curve compared to the Langevin model. Although at $\lambda_e = 0.5$, the VFE magnetization curve behaves normally, for $\lambda_e  \geqslant 1$ the VFE theory is much greater than 1, which indicates the non-physical nature of this approach in this parameter range. The LFE theory is in a good agreement with simulation data in all considered range of parameters due to the fixed logarithmic form of the free energy used for the calculation of the magnetization curve. However, in a weak field $\alpha \leqslant 1$ for $\lambda_e = 2$ some disagreement of the LFE theory with the MD results is observed.

Numerical magnetization results can be used to estimate the configurational part of the Helmholtz free energy in the following way:
\begin{eqnarray}\label{F_num}
\frac{\beta \Delta F}{N} &=& \int \limits _0 ^\alpha\left[ L(\alpha) - M \right] d\alpha + \frac{\beta \Delta F (\alpha = 0)}{N}.
\end{eqnarray}
\noindent However, the numerical value of $\Delta F (\alpha = 0)$ is unknown; therefore, only the relative increase of $\Delta F $ can be considered as
\begin{eqnarray}\label{F_alpha}
\frac{\beta \Delta F_{\alpha}}{N}  &=& \left| \frac{\beta \Delta F}{N} - \frac{\beta \Delta F (\alpha = 0)}{N}\right|.
\end{eqnarray}
\noindent Fig. \ref{fig:free_en} shows theoretical and numerical results
for the value of $\Delta F _{\alpha}$ as a function of Langevin parameter $\alpha$.
In all cases, the LFE theory looks better than the VFE theory,
which is in a good agreement with numerical data only in the weak field $\alpha\leqslant 1$ for $\lambda_e = 0.5$. For $\lambda_e = 2$, which corresponds to the strong dipole-dipole interactions in the model system, neither theory works well. It should be noted that numerical results for $\Delta F _{\alpha}$ at $\lambda_e = 2$ turned out lower than for $\lambda_e = 1.5$, although both theories increase with the growth of $\lambda_e$. The considered theoretical method is not allowed to take into account some structural magnetic affects in system, which could be the reason for the behavior of $\Delta F _{\alpha}$.

\begin{figure}[b!]
\center
\includegraphics[width=0.83\linewidth]{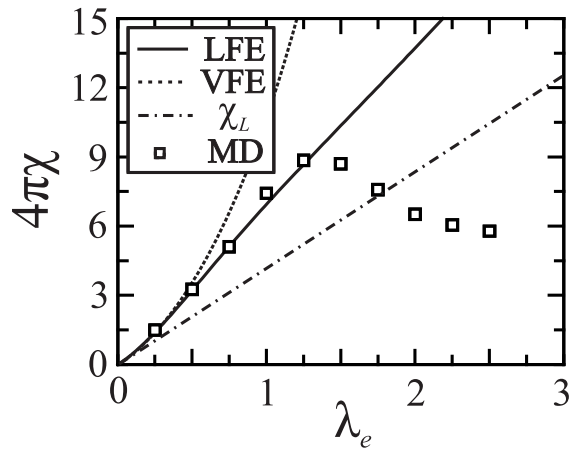}
\caption{The initial magnetic susceptibility $\chi$ as a function of dipolar coupling constant $\lambda_e$ for the SCLF. Points are from MD simulations, the solid line is Eq. (\ref{chi_log}), the dashed line is Eq. (\ref{chi_vir}), and the dash dotted line is the Langevin susceptibility.}
\label{fig:chi}
\end{figure}

The theoretical initial magnetic susceptibility $\chi$ can be expressed via the Helmholtz free energy as
\begin{eqnarray}\label{chi}
 \chi = - \frac{1}{V} \frac{\partial^2 F}{\partial H^2}\Bigg | _{H=0}.
\end{eqnarray}
In the used notation, the Langevin susceptibility looks like $\chi_L = \lambda_e / 3$. The VFE theory allows us to represent the initial magnetic susceptibility as a series over powers of $\chi_L$
\begin{eqnarray}\label{chi_vir}
 \chi = \chi_L \left[ 1 + \frac{4\pi \chi_L}{3} \left(1+0.1266 \lambda_e^2\right) \right].
\end{eqnarray}
\noindent The LFE theory gives
\begin{eqnarray}\label{chi_log}
 \chi =  \chi_L \left[ 1 + \frac{4\pi  \chi_L \left(1+0.1266 \lambda_e^2\right)}{3+4\pi \chi_L^2}  \right].
\end{eqnarray}
\noindent Comparison between the theoretical susceptibility curves and the MD simulation data is given in Fig.~\ref{fig:chi}. The Langevin susceptibility totally neglects the dipolar interactions between ferroparticles; therefore, it greatly underestimates MD results even for a system with a low dipolar coupling constant. The VFE theory has the applicability range $\lambda_e \leqslant 0.5$,
while the LFE theory allows to accurately describe the MD results up to $\lambda_e = 1.25$. After this, the nonmonotonic behavior of $\chi$ is observed in the MD simulation, although the analytical curves demonstrate ordinary increases to the initial magnetic susceptibility with growth in $\lambda_e$.
To understand the reason for the observed phenomenon, it is necessary to investigate the properties of the SCLF with the strong dipole-dipole interactions in more detail via computer simulation techniques.

\subsection{Ordering of magnetic moments in the SCLF}

\begin{figure}[t!]
%\center
\includegraphics[width=0.81\linewidth]{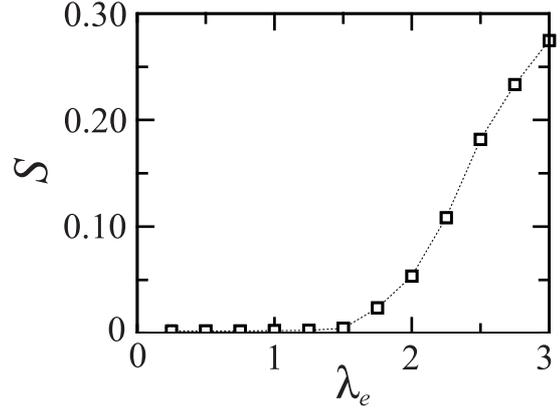}
\caption{Order parameter \emph{S} as a function of $\lambda_e$ from MD simulations.}
\label{fig:S}
\end{figure}

\begin{figure}[t!]
\center
\begin{minipage}[h]{47mm}
\center{\includegraphics[width=\linewidth]{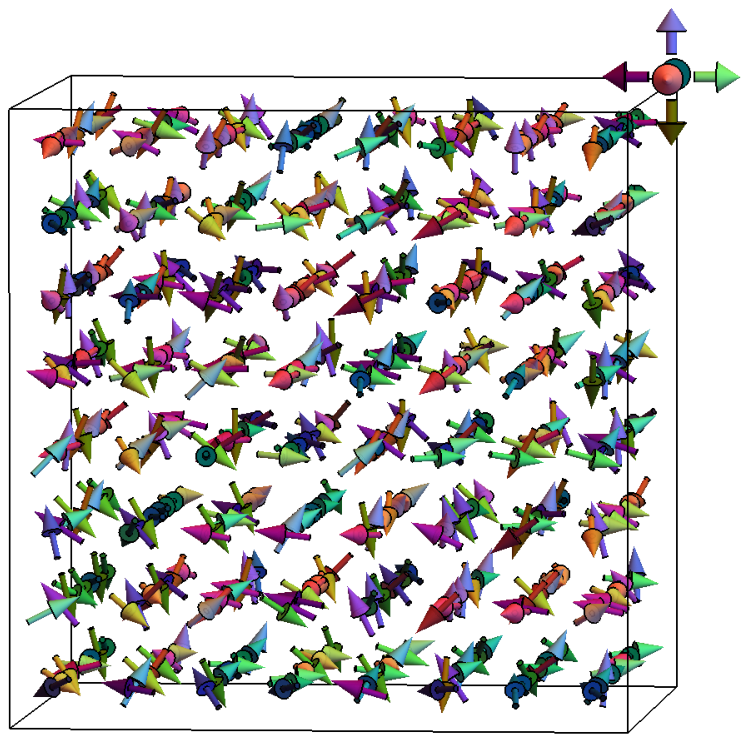} \\ (a)}
\end{minipage}\\
\begin{minipage}[h]{47mm}
\center{\includegraphics[width=\linewidth]{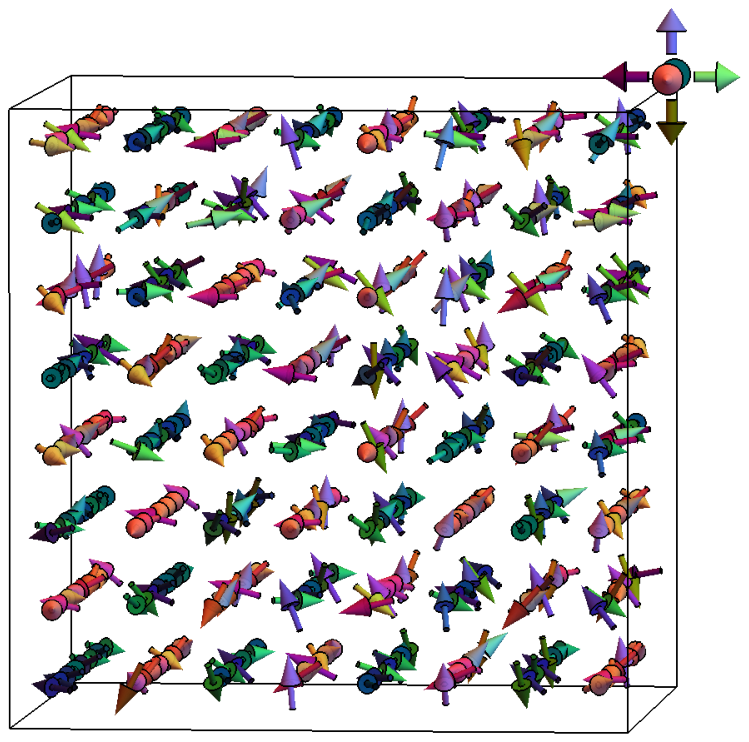} \\ (b)}
\end{minipage}\\
\begin{minipage}[h]{47mm}
\center{\includegraphics[width=\linewidth]{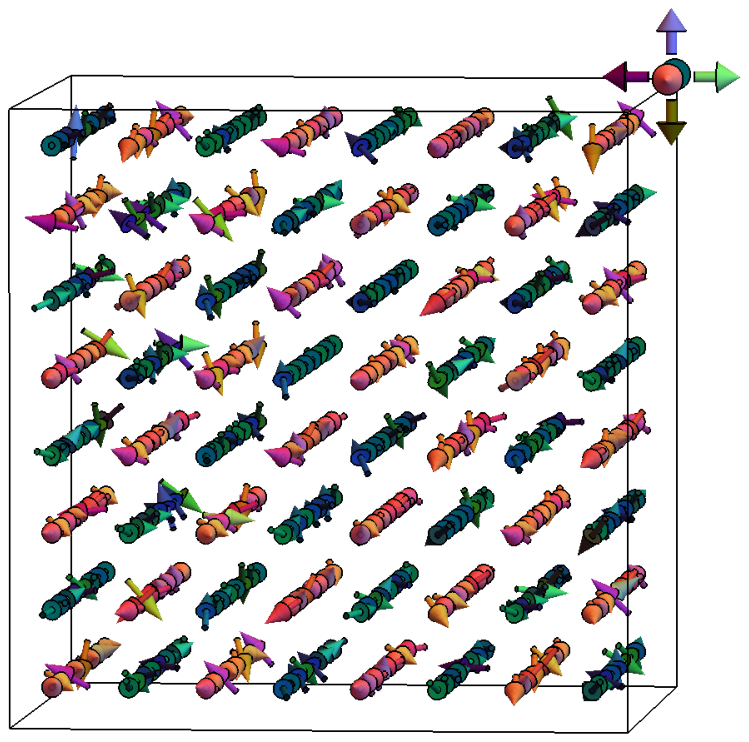} \\ (c)}
\end{minipage}
\caption{Simulation snapshots of the SCLF at $\alpha = 0$.
$\lambda_e = 1$ (a), 2 (b) and 3 (c).
Different arrow colors correspond to different orientations
of magnetic moments.}
\label{fig:order}
\end{figure}

Nonmonotonic temperature dependence of the initial susceptibility
is a common sign of phase or structural transformations in magnetic systems.
An obvious example is the ``paramagnetic -- antiferromagnetic''
phase transition in metals and alloys with a negative exchange
coupling~\cite{van1951recent,nagamiya1955antiferromagnetism}.
At high temperatures, the susceptibility of these media increases
with decreasing temperature, according to the standard Curie-Weiss law.
Below some critical temperature, known as the Curie or N{\'e}el point,
the spontaneous spin alignment takes place.
Within the simplest theoretical description, an
antiferromagnet can be modelled as a combination of two sublattices.
The spins of each sublattice are aligned with the same preferable direction (easy axis),
but the magnetisations of the two sublattices are antiparallel and cancel each other out,
so the net magnetization is zero.
If the temperature decreases below the N{\'e}el point,
the susceptibility of such an antiferromagnet along the easy axis
also decreases and vanishes at $T = 0$.
Another notable example is the initial susceptibility of
a low-concentrated DHS fluid with intensive dipole-dipole interactions.
It was shown in Ref.~\cite{kantorovich2013nonmonotonic},
both theoretically and via Monte Carlo simulations,
that with decreasing temperature DHSs in such fluid
first form linear chains, which are highly responsive
to an external field and thus lead to an anomalous increase in susceptibility.
However, as the temperature progressively decreases,
it becomes energetically favorable for a chain
to close into a ring.
Rings are magnetically inert and, with their number growing,
the fluid's susceptibility decreases.
To investigate the reasons behind the nonmonotonic
dependence of $\chi$ on $\lambda_e$ in the SCLF,
we first calculated its orientational order parameter
$S$ in zero field [Eq.~(\ref{sim_order})].
The results are shown in Fig.~\ref{fig:S}.
It is seen that at $\lambda_e \gtrsim 1.5$
(i.e., in the same range, where the susceptibility decreases
in Fig.~\ref{fig:chi}),
$S$ starts to increase rapidly,
marking the emergence of a preferable direction in
the orientation of magnetic moments.
At the same time, the overall magnetization
of the simulated system is close to zero,
regardless of how large the dipolar coupling constant is.
The specific nature of the orientational ordering in the SCLF
can be best understood from a visual examination
of the simulation snapshots given in Fig.~\ref{fig:order}.
At a large value of $\lambda_e$, all dipoles tend to align in long chains
spanning across the simulation box. 
Chains are parallel or antiparallel to some common direction, let us call it $\bm{n}$.
$\bm{n}$ always coincides with one of the lattice axes ([100], [010] or [001] in crystallographic terms).
A particular choice of $\bm{n}$ in a given simulation
run is evidently arbitrary and depends
on the initial orientation distribution.
Any chain parallel to $\bm{n}$ has four antiparallel
nearest neighbors, so that a checkerboard pattern
is observed in the plane perpendicular to $\bm{n}$.
One can also put it this way:
any particle has six nearest neighbors,
among which, two tend to adopt the most favorable
``head-to-tail'' orientation and the other four
tend to adopt the second most favorable ``side-by-side'' orientation.
Inducing a nonzero net magnetization in the system requires
breaking this energetically efficient arrangement,
which explains why the susceptibility decreases in the ordered state.
Actually, it is natural to expect that the susceptibilities along ($\chi_{||}$)
and perpendicular ($\chi_{\perp}$) to the direction $\bm{n}$ are not identical~\cite{nagamiya1955antiferromagnetism,landau1984electrodynamics}.
%We have not yet tested this hypothesis.
If this is the case, then the quantity $\chi$
calculated via Eq.~(\ref{sim_susc}) and presented in Fig.~\ref{fig:chi}
simply has the meaning of an average value, $\chi = \chi_{||}/3 + 2\chi_{\perp}/3$.

Our simulation predictions for the magnetic ordering in the SCLF
are in full agreement with the well-known results of
Refs.~\cite{luttinger1946theory,kretschmer1979ordering},
where it was shown that the ground state of a simple cubic lattice
of dipoles in a zero field is a system of ``ferromagnetic rows (...) arranged
antiferromagnetically in the plane perpendicular to it''.

It should be emphasized that the theory developed here describes correctly the properties of the model system in the area of small and moderate values of ferroparticles' concentration and the intensity of the dipole-dipole interaction. However, new theory does not allow to predict system structuring for large lambda values. This is due to the limitation in the theory up to the second virial coefficient and the third power of the effective dipolar coupling constant. Increasing the number of terms in the virial series can expand the scope of the theory. Nevertheless, using the virial expansion approach it is quite difficult to obtain the theory that is valid in the field of strong dipole-dipole interactions. In this case, an alternative variation method \cite{Kolesnikov2019} can be used to study the behavior of the system at the strong coupling regime when the magnetic moments of particles are ordered antiferromagnetically.

\section{Conclusion}\label{sec:cncld}
A new thermodynamic theory of interacting single-domain ferroparticles embedded in a simple cubic lattice has been derived from the rigorous methods of statistical physics. Accounting for pairwise dipole-dipole interactions, an explicit analytical expression for the Helmholtz free energy was obtained in the virial form (VFE-theory) Eq. (\ref{F_solid_final}) and the logarithmic form (LFE-theory) Eq. (\ref{F_solid_log}). This level of accuracy corresponds to taking into account only the second virial coefficient in the classical virial expansion, well known from textbooks and developed for a liquid (ensemble of moving particles).

The new theory has been compared against MD computer simulation results. A comparison has been made in terms of the Helmholtz free energy, the static magnetization, and the initial magnetic susceptibility for the effective dipolar coupling constants $\lambda_e \leqslant 2$. The VFE theory greatly overestimates the MD results; moreover, at rather high values of the effective dipolar coupling constant ($\lambda_e \geqslant 1$), the limitation with only a certain number of terms of $\lambda_e$ and the use of the finite polynomial instead of an infinite virial series results in nonphysical dependence of the thermodynamic functions on the coupling constant and the Langevin parameter (for example, Fig.~\ref{fig:mag} (b)-(d)). To overcome this problem, we used the LFE theory, which turns out to be in good agreement with simulation data. For $\lambda_e \leqslant 1.5$, the agreement between the LFE theory and simulation results is excellent, while for $\lambda_e = 2$, the deviations are due to the truncation of the Helmholtz free energy expansion in $\lambda_e$. From a practical point of view, the LFE theory is extremely simple: any thermodynamic function can be determined using standard relations from the Helmholtz free energy expression (\ref{F_solid_log}) for an ensemble of interacting single-domain ferroparticles embedded in a simple cubic lattice.

The features introduced by the immobility of particles into the thermodynamic properties of the system have been analyzed. The difference between the Helmholtz free energy of mobile dipole hard spheres (DHS) and the ones embedded in a cubic lattice (SCLF) increases with increasing dipole-dipole interaction and the intensity of the external field. Using the LFE theory and the MD simulations, it has been found that at low intensities of the dipole-dipole interaction ($\lambda_e \lesssim 0.5$) the magnetization of DHS and SCLF is the same. For strong and moderate values of the coupling constant $0.5 \lesssim \lambda_e \lesssim 2$ and the weak magnetic field ($\alpha \lesssim 0.5 $), the magnetization of the DHS system is higher than the magnetization of SCLF, while the opposite tendency is observed in stronger fields. The reasons for this behavior are quite clear. In a weak magnetic field, the dipole-dipole interaction plays a main role in the system and leads to the formation of dimers in DHSs, that increases the magnetization. In the SCLF system, where particles do not have translational degrees of freedom, the appearance of dimers is impossible; therefore, the magnetization of the SCLF system is less than in a DHS fluid. In strong and moderate magnetic fields with strong and moderate dipole-dipole interactions, a competition between the energy of interparticle dipole-dipole interaction and the field - magnetic moment interaction energy occurs. Balance is achieved by the formation of correlation structures in which particle magnetic moments can compensate for each other. In the SCLF, the correlation structures begin to form at $\lambda_e > 1.5$ even in the absence of a field (Fig. \ref{fig:order}). In the DHS fluid, particles can be located at distances shorter than the lattice size $a$ up to the particle contact $\sigma$, that is,  the intensity of interparticle interaction in DHS fluid is determined by the parameter $\lambda = \lambda_e a ^ 3 / \sigma ^ 3$, which exceeds the value of $\lambda_e$. This leads to the fact that the formation of the correlation structure with compensating magnetic moments in the DHS fluid occurs at much lower values of $\lambda_e$.

Using the methods of MD computer simulation, the internal  structure of SCLF  has been analyzed for large values of $\lambda_e$. It has been shown that for $\lambda_e > 2$, an antiferromagnetic order appears in the system.

Overall, the LFE theory provides a reliable means of predicting thermodynamic properties of the SCLF. The current investigation represents essential information in the design and synthesis of new functional materials.

\section*{Acknowledgements}\label{Ack}
The reported study was funded by RFBR, project number 20-02-00358. AYS and EAE acknowledged Prof. Philip Camp for his advice and useful discussions.

\bibliography{ref}

\end{document}